\documentclass[twocolumn,tighten]{aastex63}

\usepackage[T1]{fontenc}

\submitjournal{ApJ}

\shortauthors{Zaritsky et al.}

\begin{document}

\title{A Lower Limit on the Mass of Our Galaxy from the H3 Survey}

\correspondingauthor{Dennis Zaritsky}
\email{dennis.zaritsky@gmail.com}

\author{Dennis Zaritsky}
\affiliation{Steward Observatory, University of Arizona, 933 North Cherry Avenue, Tucson, AZ 85721-0065, USA}
\author{Charlie Conroy}
\affiliation{Center for Astrophysics | Harvard \& Smithsonian, 60 Garden Street, Cambridge, MA 02138, USA}
\author{Huanian Zhang}
\affiliation{Steward Observatory, University of Arizona, 933 North Cherry Avenue, Tucson, AZ 85721-0065, USA}
\author{Ana Bonaca}
\affiliation{Center for Astrophysics | Harvard \& Smithsonian, 60 Garden Street, Cambridge, MA 02138, USA}
\author{Nelson Caldwell}
\affiliation{Center for Astrophysics | Harvard \& Smithsonian, 60 Garden Street, Cambridge, MA 02138, USA}
\author{Phillip A. Cargile}
\affiliation{Center for Astrophysics | Harvard \& Smithsonian, 60 Garden Street, Cambridge, MA 02138, USA}
\author{Benjamin D. Johnson}
\affiliation{Center for Astrophysics | Harvard \& Smithsonian, 60 Garden Street, Cambridge, MA 02138, USA}
\author{Rohan P. Naidu}
\affiliation{Center for Astrophysics | Harvard \& Smithsonian, 60 Garden Street, Cambridge, MA 02138, USA}

\begin{abstract}
The timing argument provides a lower limit on the mass of the Milky Way. We find, using a sample of 32 stars at $R > 60$ kpc drawn from the H3 Spectroscopic Survey and mock catalogs created from published numerical simulations, that M$_{200} > 0.91\times 10^{12}$ M$_\odot$ with 90\% confidence. We recommend using this limit to refine the allowed prior mass range in more complex and sophisticated statistical treatments of Milky Way dynamics. The use of such a prior would have significantly reduced many previously published uncertainty ranges.
Our analysis suggests that the most likely value of M$_{200}$ is $\sim 1.4 \times 10^{12}$ M$_\odot$, but establishing this as the Milky Way mass requires a larger sample of outer halo stars and a more complete analysis of the inner halo stars in H3. The imminent growth in the sample of outer halo stars due to ongoing and planned surveys will make this possible.
\end{abstract}

\keywords{Galaxy: fundamental parameters --- Galaxy: halo --- Galaxy: kinematics and dynamics
}

\section{Introduction}

The mass of our own galaxy would seem to be one of a modest set of fundamental parameters that is both necessary for a complete understanding of our universe and relatively straightforward to measure. However, convergence on a precise final answer has proven difficult.
\cite{zar89} presented three mass estimates, two of which are based on a statistical treatment of satellite galaxy radial velocities and one of which is a lower limit based on the properties of a single, extreme satellite (Leo I):
($0.93^{+0.41}_{-0.12},1.25^{+0.84}_{-0.32},  1.3^{+0.33}_{-0.33})\times 10^{12} {\rm M}_\odot$.
Exemplar, state-of-the-art studies conclude that the Galaxy's mass is
 ($0.85^{+0.23}_{-0.26},
0.96^{+0.29}_{-0.28}, 1.29^{+0.37}_{-0.37})\times 10^{12} {\rm M}_\odot$ based on satellite dynamics and the stellar escape velocity \citep{patel18,grand2019} .
The casual observer may be somewhat underwhelmed by 30 years of ``progress''. 

This cursory comparison is misleading.
Studies in the intervening years identified areas ripe for improvement in earlier analyses \citep[e.g.,][]{li, patel18}, utilized the ever-increasing amount of available data \citep[e.g.,][]{watkins}, most importantly by adding proper motions \citep[e.g.,][]{bk}, and developed the cosmological framework in which we construct our models \citep[e.g.,][]{barber,carlesi,patel17}. The surprise is that better data in combination with more sophisticated models did not manifestly result in evident quantitative improvements. Derived parameter uncertainties grew or remained at best constant even as the quantity and quality of the data improved because the increasingly sophisticated models contain more degrees of freedom. The apparent stagnation is therefore partly an illusion. Improvements over the years have implicitly tested for systematic biases in previous, oversimplified and underconstrained models. However, as happens sometimes, there were no such biases of a magnitude above and beyond the quoted precision of the earlier measurements.
 This result should be both disappointing and reassuring to  practitioners in the field.

How to proceed? There is a tension between using increasingly realistic models and the larger parameter space these models access. One way to mitigate this tension is to develop more constraining parameter priors. Lower mass limits, such as provided by an escape velocity measurement or the timing argument, could possibly help, but have historically been discounted when in conflict with other results \citep[e.g.,][]{xue, gibbons}. The rejection of this additional information was not unreasonable given that the largest (most constraining) lower mass limit came from the study of the single satellite galaxy Leo I \citep{zar89,li}, which might not be on a simple orbit \citep{sales,mateo}\footnote{An alternative lower limit that is also highly constraining comes from the Local Group timing argument, but that measures the sum of the Milky Way and M 31 masses. In this case uncertainty is introduced by the partitioning of the calculated mass.}. Subsequent work, including new proper motion measurements and a careful theoretical treatment, concluded that Leo I is indeed bound to the Galaxy and likely to be a strong constraint \citep{bk}. However, any mass limit from Leo I is still susceptible to the usual 
concern of relying too heavily on the properties of a single object. 

Our aim is to reinforce a meaningful lower mass limit.
In doing this we have 
two key points of emphasis. First, we aim to use the most distant dynamical tracers available to measure the total Galactic mass without having to extrapolate too much in radius. 
Meeting this goal can be challenging because tracer particles are generally available in quantity only to distances $<$ 50 kpc \citep[e.g.,][]{williams} and the Milky Way's virial radius, is likely to be 250 kpc or larger.
Second, we want to obtain a conservative, but stringent inference using a straightforward and simple analysis.

The two principal methods for determining a Galactic lower mass limit are based on the escape velocity and the timing argument\footnote{A non-dynamical lower limit can be formulated using the cosmological baryon fraction \citep{zar17}, but the uncertainties are large until the circumgalactic gas mass is measured more precisely.}. The escape velocity argument is highly intuitive: the Galaxy needs to have sufficient mass to retain the fastest moving stars that are still found  near the Galaxy. Several treatments of the escape velocity measurement have been published recently \citep{piffl,williams,deason,grand2019}. While limits can be obtained from the fastest moving stars in any sample, the most meaningful limits come from large samples of stars where the high velocity tail truly includes stars moving at nearly the escape velocity. Such studies are necessarily carried out with samples of stars at modest Galactic radii of, at best, a few tens of kpc. 
Because of this limitation, evaluating the total Galactic mass requires a model of the Galactic potential and the phase space distribution of the stars. The uncertainties arising from the model choice can then be dominant (for example, see discussion by \cite{deason} comparing their inference to that of \cite{piffl}, which differ despite similar direct measurements.)

The second method that can provide a
lower mass limit is the timing argument \citep{kw,sandage}. The timing argument posits that all matter initially expands away from the future galaxy due to universal expansion and that the subsequent gravitational pull of the galaxy eventually recalls some of that matter. In the simplest version of this scenario, the tracer particles, galaxies or stars, follow a radial orbit around a point mass. In a slightly more complex version, orbit crossing leads to the behavior seen in secondary infall models without angular momentum \citep{fillmore,bertschinger} or with angular momentum \citep{white}.
More complete treatments are currently being investigated to define the splashback radius and study the growth of structure \cite[e.g.,][]{diemer}.

In the simplest version, the system mass can be solved for analytically if the current separation and relative velocity between the Galaxy and a tracer particle are known. One does not need to know the phase space distribution of the population of tracer particles.
The advantage provided by the timing argument comes from using one additional piece of information, namely the constrained orbital time (which is the age of the Universe). Because this analysis does not require us to assume either a velocity distribution function or radial tracer density profile, we do not need detailed knowledge of the accretion history of the tracer particles or the assumption of dynamical relaxation. 

The timing argument is an exceedingly elegant and robust treatment, particularly given the now well-measured age of the universe. There are a few, limited caveats. First, the standard solution is for a radial orbit. If the tracer particle has non-zero tangential velocity then the actual mass can be significantly larger than that inferred \citep[see][for how the additional tangential velocity component of Leo I as measured with {\sl HST} affected the inferred Galactic mass]{bk}. Second, the Galaxy is modeled as a point mass. Neglecting both its extent and its growth over time result in the classical timing argument underestimating the current mass within the orbit of the tracer particle. 
Finally, there is the possibility of an invalid result if the tracer particle has a large peculiar velocity resulting from interactions between the tracer particle and other mass concentrations. This, for example, was one of the principal concerns in the application of the timing argument to Leo I \citep{sales,mateo}. If Leo I has been flung toward the Milky Way due to an encounter with a third galaxy, then its current position and velocity do not reflect those of the assumed cosmologically-motivated, radial orbit about the Milky Way. 
All of these concerns are addressed when using numerical simulations both to calibrate the estimator and to determine the likelihood of catastrophic failures. Catastrophic failures can be mitigated by repeating the analysis using multiple, independent tracer particles.

Numerical simulations confirm 
that the simple timing argument systematically underestimates the system mass and that statistical corrections are necessary to accurately recover the true mass \citep{kroeker, li}.
For the Local Group timing argument, using the orbit of M 31 and the Milky Way, that statistical correction requires increasing the result obtained from the application of the simple timing argument by about a factor of 1.7. Those studies also confirm that the scatter between the estimated and true mass can be large when using only M 31 and the Galaxy or Leo I and the Galaxy.

We present the application of the timing argument to a sample of outer halo ($R > 60$ kpc) stars from the H3 Spectroscopic Survey \citep{conroy}. 
We discuss the sample of stars in \S2, the application and results from the timing argument to both the models and data in \S3, discuss the results in \S4, including an illustration of how constraining a robust lower mass limit can be on previous results, and summarize our conclusions in \S5.

\section{The Sample}

\begin{figure}[htbp]
\begin{center}
\includegraphics[width = 0.48 \textwidth]{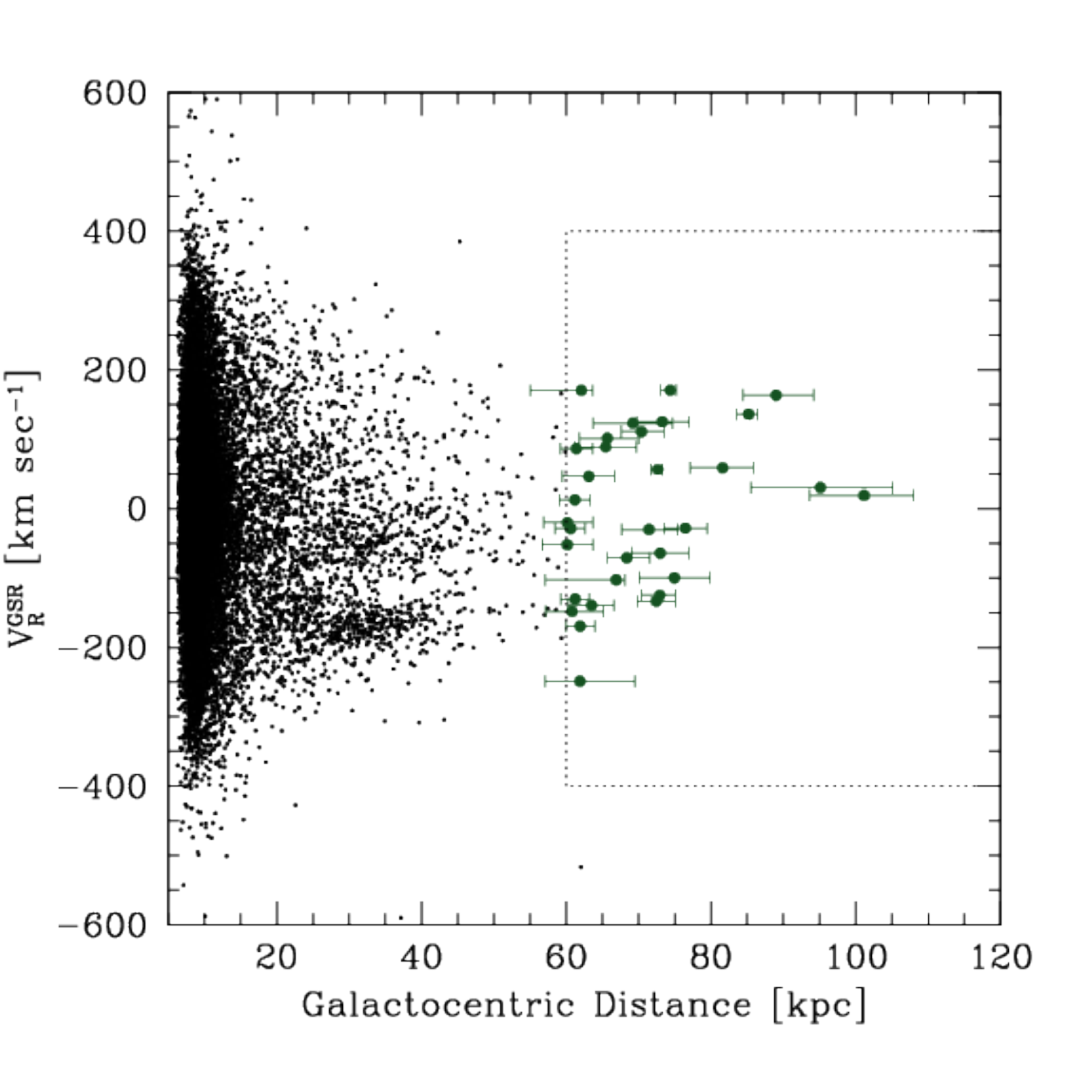}
\end{center}
\caption{The H3 sample and outer halo stars. From the full sample we select a sample of 32 outer halo stars ($R > 60$ kpc, $-400 \le {\rm v}_{\rm R}^{\rm GSR} \le 400$ km sec$^{-1}$) for the timing argument analysis, as shown highlighted within the dotted selection box. }
\label{fig:data}
\end{figure}

H3 is a sparsely sampled survey of $\sim$ 15,000 square degrees of sky that will provide stellar
parameters and spectrophotometric distances for $\approx$ 200,000 stars to $r = 18$  \citep{conroy}.
The procedure used for the parameter determinations and distance estimation is developed and presented by \cite{Cargile2019}. From the currently available set of observed and analyzed stars, about 80,000, we select those with fitting flag set to 0 (denoting fitting was problem-free), spectral signal-to-noise ratio (SNR) per pixel $>$ 3, a stellar rotational velocity $<$ 5 km sec$^{-1}$, an effective temperature $<$ 7000 K, and a tangential velocity, ${\rm v}_{\rm T}$, that is $ < 1000$ km sec$^{-1}$. These criteria are all meant to remove stars with problematic and unphysical parameter determinations and were arrived at after a careful examination of numerous fits. 
From that sample, we then select those stars with Galactocentric distance $>$ 60 kpc (Figure \ref{fig:data}) to focus on stars that are likely to have completed only a few Galactic orbits at most.

In Figure \ref{fig:data} we highlight the 32 outer halo stars within a selection box that adds the criteria that the radial velocity in the Galactic Standard of Rest, ${\rm v}_{\rm R}^{{\rm GSR}}$, satisfies $|{\rm v}_{\rm R}^{\rm GSR}| < 400$ km sec$^{-1}$. 
This selection removes the one star at $R > 60$ kpc that clearly does not belong in our sample (with ${\rm v}_{\rm R}^{\rm GSR} < -500$ km sec$^{-1}$). Whether this star represents a particularly catastrophic parameter determination or a physically compelling outlier, is unknown and merits follow-up, but in neither case is this star suitable for our mass estimation. A preliminary investigation of the model fit for this one star suggests that this system has a large residual flux in the IR and is, therefore, perhaps an unresolved binary star for which the parameter estimation failed. The radial velocity errors in general are on the order of a few km sec$^{-1}$, so the distance errors dominate in our calculations and are plotted in the figure. 
The distribution of stars follows the apparent caustics that can be visually defined at $R < 60$ kpc. Substructure is clearly evident at $R < 50$ kpc and illustrates the difficulty faced by any statistical analysis of the velocities where one needs to model the phase space distribution.

\section{The Timing Argument}

The basic timing argument is simple in its application. Using the equations provided by \cite{sandage}, we find the smallest orbital phase, $\theta$, that solves the equations relating distance, velocity, and orbital time for a selected star. The orbital time is taken to be the age of the universe, 13.75 Gyr \citep{wmap}. Once we solve for the orbital phase, we calculate the mass.

\subsection{Calibration Using Simulations}

\begin{figure}[htbp]
\includegraphics[width = 0.48 \textwidth]{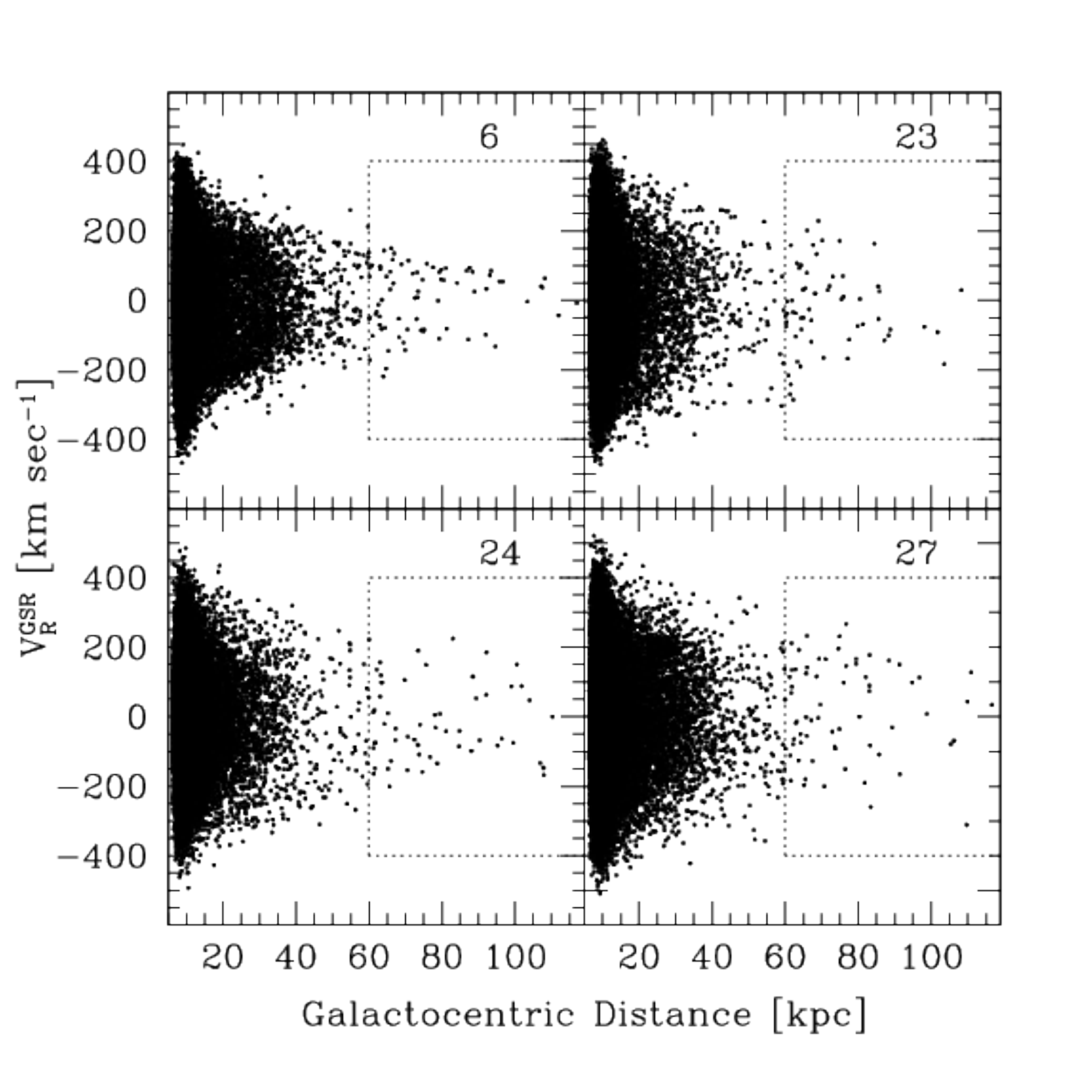}
\caption{Mock H3 catalogs created from Auriga models (6, 23, 24, and 27) from \cite{grand2017}. The outer halo star selection box from Figure \ref{fig:data} is also shown. The model number is given in the upper right of each panel. Each model contains roughly twice as many stars in the selection box than H3 because we apply the additional data quality and model fitting criteria to the actual data.}
\label{fig:sims}
\end{figure}

We construct H3 mock catalogs from the Auriga suite of simulations \citep{grand2017}.  Specifically, we have catalogs that mock the H3 selection, including a 10\% distance uncertainty, for their model numbers 6,16,21,23,24, and 27. We construct the plots corresponding to our Figure \ref{fig:data}, applying all of the same selection criteria except those aimed at removing stars with poor spectral fits. We find that models 16 and 21 have dominant substructures visible at large radii ($R > 60$ kpc) that are qualitatively different from what we find in H3 (but perhaps similar to what is seen in H3 at smaller radii, between 25 and 40 kpc). The remainder of this discussion uses only the four models that do not show such structure (see Figure \ref{fig:sims}) and are, at least visually, plausible matches to the data. 

\begin{deluxetable}{rrr}
\label{tab:auriga}
\tablewidth{0pt}
\caption{Auriga Model Parameters}
\tablehead{
\colhead{Auriga Run}  &
\colhead{$M_{200}$}& \colhead{$R_{200}$}\\ 
&\colhead {[$10^{12} {\rm M}_\odot$]} & \colhead {[kpc]}}
        \startdata
         6&1.04 &214 \\
         23&1.58&245 \\
         24&1.49 &241\\
         27&1.75&254 \\
         \enddata
\end{deluxetable}

We are interested in using timing argument calculations for stars at 60 kpc $< R < 120$ kpc to recover M$_{200}$, where M$_{200}$
is the mass contained inside a radius within which the mean density equals 200 times the critical density. When we apply the timing argument calculation to all of the star particles in each of these four models and then scale the recovered timing mass and the calculated apocenter\footnote{We utilize the calculated apocenter distances using the timing argument equations rather than the apocenter distances presented in the H3 catalog, which are calculated in a completely independent manner. We do this to avoid introducing different model assumptions into our current analysis and discussion.} of each star particle by  the corresponding M$_{200}$ and R$_{200}$ of each model, we find the relation between derived mass and apocenter shown in the upper panel of Figure \ref{fig:apocenters}. 

As expected, the timing mass significantly underestimates M$_{200}$ in all cases. The closest any estimate gets to the true value in any of the four models is less than half of the true value. The models show very little scatter in the estimates along the sequence, demonstrating that strong orbital perturbations are not significantly affecting the estimates. Because the four models together have roughly eight times as many stars as we have in the H3 sample (257 vs. 32), the likelihood of a star in H3 resulting in an estimate closer to M$_{200}$ than what we find in the models is small ($\sim 12$\%). Therefore, 
the results from analyzing the models suggest that one can take the star with the largest timing mass estimate in H3, multiply that result by 2.3 (1/0.43) to estimate M$_{200}$, and still have a conservative estimate of M$_{200}$. 

\subsection{Application to the H3 Outer Halo Sample}

Among the H3 sample, the largest resulting timing mass comes from H3 star 117408280 ($\alpha = 355.479908, \delta = 15.82972)$ and is $0.49 \times 10^{12}$ M$_\odot$. Using the calibration factor of 2.3, yields a timing mass estimate for M$_{200}$ of $1.1 \times 10^{12}$ M$_\odot$. Simulating the effect of the distance uncertainty by drawing 1000 realizations of the distance and re-evaluating the resulting mass estimate, we find that with 90\% confidence M$_{200} > 0.97 \times 10^{12}$ M$_\odot$.
This star has a negative radial velocity, eliminating any concern that this star happens to be an unbound, hypervelocity star \citep{brown}. 

\begin{figure}[t]
\includegraphics[width = 0.48 \textwidth]{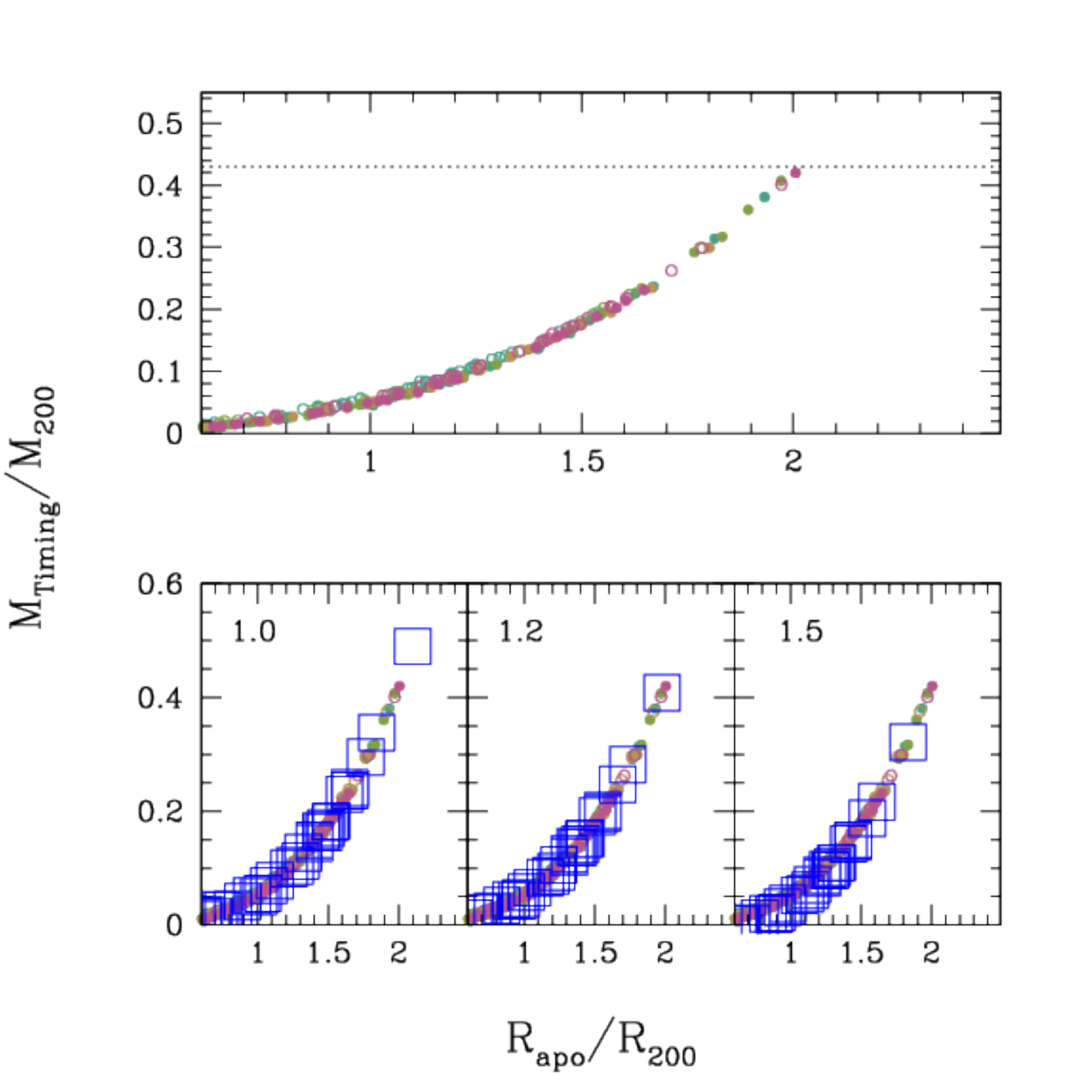}
\caption{Distribution of scaled timing argument results. We scale the derived mass by M$_{200}$ for each of the outer halo star particles in each of the four simulations and the calculated apocenter distances by R$_{200}$ in the upper panel. The different models are represented by different colors and inbound and outbound stars by closed and open symbols, respectively. All estimates are less than half of the true value regardless of the model and whether the inward or outward moving stars are used. The dotted horizontal line represents M$_{\rm Timing}$/M$_{200} = 0.43$ and lies above all estimates.
In the lower panels, we reprise the simulation results in all panels and present the scaled values for the H3 outer halo stars, large open squares, for three different adopted values of M$_{200}$ ($10^{12}$, $1.2\times 10^{12}$ and $1.5\times 10^{12}$ M$_\odot$) in each of the three panels, as labeled. There are roughly a factor of eight more tracer particles in the combined simulations, so we do not expect the H3 sample to sample the extreme upper end of the distribution better than the simulations. This consideration and visual examination of the Figure suggest that M$_{200}$ for the Milky Way is $> 10^{12}$ M$_\odot$.}
\label{fig:apocenters}
\end{figure}

The weakness of this argument is that it relies on identifying the largest, most constraining timing mass estimate. Again, as with Leo I, we are at the mercy of a single object. Therefore, we now approach the problem in a complementary manner by looking at the distribution of timing mass estimates. We
compare the distribution of M$_{Timing}/$M$_{200}$ for the H3 data to that drawn from the four simulations.  Of course, we do not yet know M$_{200}$ for the Galaxy. However, if we choose an acceptable value of M$_{200}$ then the resulting H3 distribution should be statistically consistent with those produced from the mock catalogs. We visualize this approach in the bottom panels of Figure \ref{fig:apocenters}, where we have rescaled the H3 results using three hypothetical values of M$_{200}$ for the Milky Way ($10^{12}, 1.2\times 10^{12},$ and $1.5\times 10^{12}$ M$_\odot$).
To calculate the radial rescaling, we use the relationship between M$_{200}$ and R$_{200}$ given by the Auriga simulations (Appendix A). The differences are most evident at the largest values of M$_{Timing}/$M$_{200}$, but the entire distribution of points slides along the locus and this behavior is what we use to obtain our mass estimate.

We apply the Kolmogorov-Smirnov (KS) test to assess which rescaled H3 M$_{\rm Timing}/$M$_{200}$ distributions can be discriminated from that of the mock catalogs. If we can reject with $>$ 90\% confidence the hypothesis that the resulting M$_{\rm Timing}/$M$_{200}$ distribution is drawn from the same parent sample as the distribution obtained from the mock catalogs, then we reject the adopted M$_{200}$ as valid for the Milky Way. In this way, we conclude that M$_{200} > 0.91 \times 10^{12}$ M$_{\odot}$ (and that M$_{200} < 2.13 \times 10^{12}$ M$_\odot$). The KS test, by construction, tends to focus on deviations in the middle of the distribution rather than at the extremes, making this a valued complement to our prior estimate that used only the most extreme star. Indeed, this general behavior is also the case here, where the maximum deviation in the cumulative distributions between data and mocks for our lower limit of M$_{200} = 0.91 \times 10^{12}$  M$_{\odot}$ occurs halfway through the ranked values of M$_{\rm Timing}/$M$_{200}$ for the H3 stars. 

The weakness of this argument is that because we are relying on the distribution of a sample of tracers, such as is done in the escape velocity estimation, we are assuming that the model is a fair representation of both the tracer distribution in phase space and the underlying potential. We have already seen that substructure can vary  significantly from model to model. We mitigate this concern by combining four different models, hoping to average over any undetected substructure at $R > 60$ kpc, but even so, an underlying systematic difference in the modeling can skew the result. 

It is highly reassuring that both approaches, one based on the most extreme outer halo star in the sample and the other based on something closer to the typical outer halo star, produce nearly identical limits. We adopt the smaller of the two limits as the limit that we quote (M$_{200} > 0.91 \times 10^{12}$ M$_\odot$) but note that the preferred values ($1.1 \times 10^{12}$ M$_\odot$ from the single, extreme star and $1.38 \times 10^{12}$ M$_\odot$ from the most likely scaling value) provide further guidance on what the Milky Way mass might actually be.

\section{Discussion}

These results will strengthen as the number of independent probes increases.
Our mass limit is consistent with that drawn previously from Leo I \citep{zar89,bk}, lending further credence to both our current limit and that from Leo I, and more data are imminent. First, H3 itself when complete will be three times the size of the current sample. Second, other large halo star surveys are  ongoing or about to start \citep{majewski,martell,deng,dalton,guiglion} from which we expect to gain more examples of such stars.  It will be increasingly more difficult to dismiss a lower mass limit of $\sim 10^{12}$ M$_\odot$.

\begin{figure}[t!]
\begin{center}
\includegraphics[width = 0.48 \textwidth]{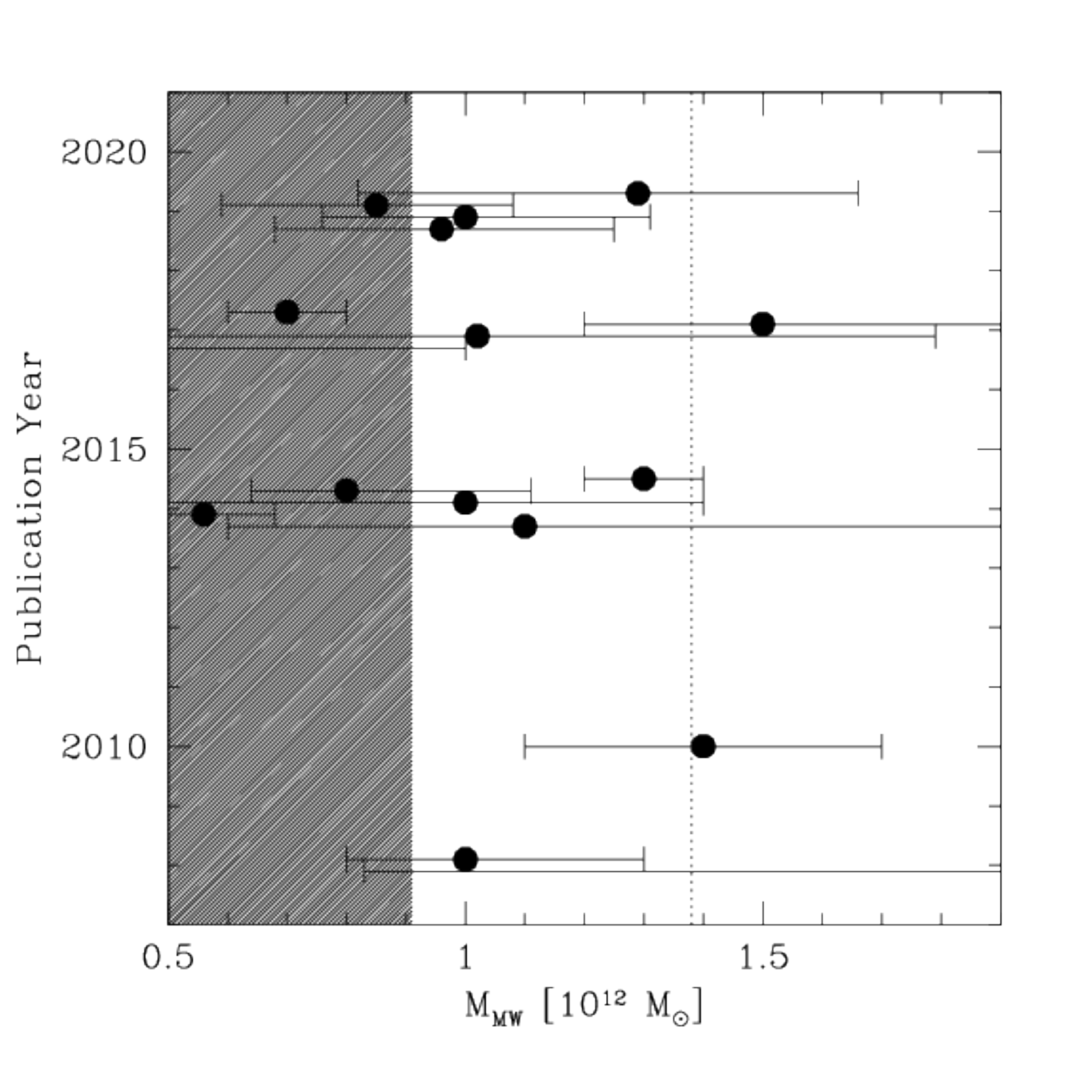}
\end{center}
\caption{A sample of literature measurement of the Milky Way mass. The shaded region illustrates the region excluded by our 90\% confidence lower limit. Only two previous measurements are strongly argued against, but for most of the measurement our lower limit reduces the allowed range of values and so provides a useful prior. The data shown represent results (from oldest to most recent): \cite{li,xue,watkins,barber,gibbons,cautun,kafle,piffl,dierickx,patel17,fragione,carlesi,patel18,deason,watkins} (\cite{patel18} is represented twice, as the fourth and second to last).  The dashed vertical line represents our preferred value derived from the comparison of actual vs. model timing mass estimates.}
\label{fig:lit}
\end{figure}

\begin{figure}[t!]
\begin{center}
\includegraphics[width = 0.48 \textwidth]{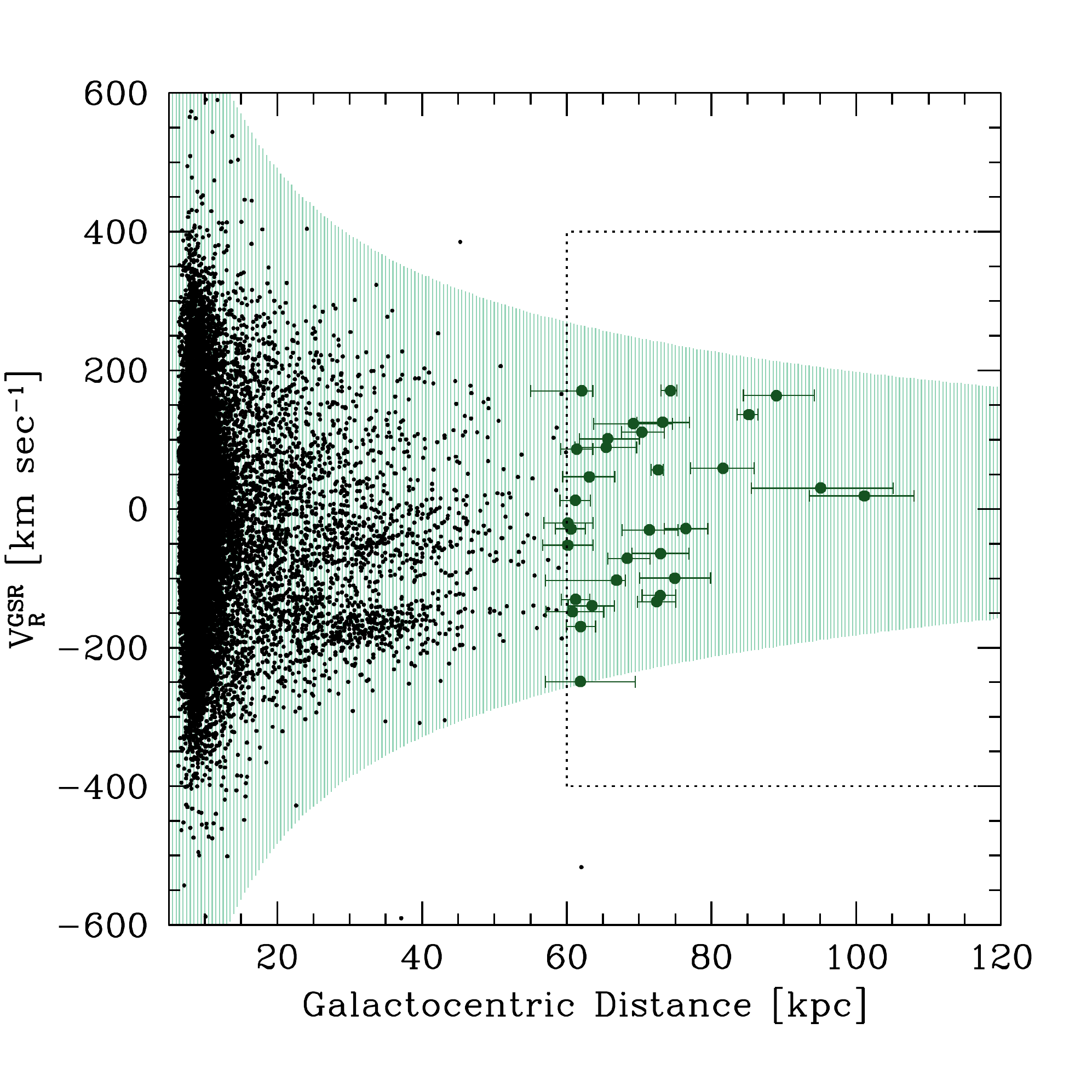}
\end{center}
\caption{The timing argument and M$_{200} = 1.1\times 10^{12}$ M$_\odot$. We compare the distribution of stars in this phase space, reproducing Figure \ref{fig:data} to the region that produces timing mass estimates equal to or smaller than that produced by our most constraining H3 star at $R > 60$ kpc (shaded region). The shaded region traces faithfully the distribution of stars at $R < 60$ kpc and leaves little doubt that the few stars outside the shaded region are some type of interloper rather than indicative of a much larger Milky Way mass. }
\label{fig:timing_orbits}
\end{figure}

\subsection{How Constraining Is Our Lower Mass Limit?}

In Figure \ref{fig:lit}, we compare a selection of recent measurements (publication since 2008) of the Milky Way mass to our 90\% confidence lower mass limit of $0.91\times10^{12}$ M$_\odot$.
Most of the published masses refer to either M$_{200}$ or the virial mass. Only in one of these cases is the mass quoted within a fixed physical radius of 300 kpc. All of these are thought to be roughly interchangeable given the other outstanding uncertainties.

Each of the published analyses, even if they incorporate data from previous work, is independent. We argue that once provided with a robust, non-trivial lower limit each study could significantly benefit by limiting the mass range over which they consider models. In many cases in the figure, about half of the internally allowed Milky Way mass range is inconsistent with the lower limit we provide.

Two results whose uncertainty ranges lie entirely or nearly entirely within our excluded zone in Figure \ref{fig:lit}
are those of \cite{gibbons} and \cite{dierickx}. Both of these are based on an analysis of the Sagittarius dwarf, suggesting that perhaps there is something that is not well understood about this system. A third recent analysis of this system concludes that the enclosed mass within 100 kpc is $\sim$ $7 \times 10^{11}$ M$_\odot$ \citep{fardal}, which is in tension with the previous estimates and more in line with our estimates.

\subsection{Could Our Limit be a Significant Underestimate?}

We associated the star with the largest timing mass estimate with the upper end of the modeled timing mass estimates to derive a mass limit. It is possible that this star does not represent the upper end of the distribution of Milky Way stars and results in a significant underestimate of the total mass. The complementary limit obtained by using the full distribution of timing mass estimates argues against such a premise, but we noted that this analysis has its own potential problems, namely its dependence on models that might not be producing an accurate match to the phase-space distribution of the dynamical tracers even if they are accurately representing the gravitational potential.

We previously rejected one star at very large negative velocities from the sample and we have not yet considered stars with $R< 60$ kpc. Are there similar stars at $R< 60$ kpc that support an argument for a significantly larger mass estimate?
In Figure \ref{fig:timing_orbits} we reprise our presentation of the distribution of H3 stars and highlight, in the shaded region, where stars need to be to result in timing mass estimates that are less than or equal to that of our most constraining star at $R > 60$ kpc. We find only three stars in the Figure that are outside the shaded region (there are a few more beyond the boundaries of the plot but those are very likely to be spurious). We conclude that there is no indication, even when considering stars at $R<60$ kpc, of a convincing population of stars that would support a mass that is  significantly larger than that resulting from the most extreme star in our sample at $R>60$. Instead, we find a number of stars near the boundary of the shaded region at both positive and negative ${\rm v}_{\rm R}^{\rm GSR}$, providing support for the value of the timing mass estimate derived previously using the one star. 

This figure raises the question of whether we should be fitting to data at all radii, rather than just to outer halo stars. There are, however, a few complications in working with stars at smaller radii. First, the steepness of the caustic curve at small radius means that the mass estimation is much more sensitive to distance errors than it is at the larger distances we have worked at. Additionally, the radial velocities are also more dependent on the proper motion measurements than for stars at large distance, where the line-of-sight velocity and the radial velocity in the Galactic frame are closer to being parallel. Lastly, working at small radii there are greater concerns regarding the impact of perturbations. All of this does not lead us to conclude that these data are not useful, but rather that they require a more complete treatment, including perhaps a closer examination of numerical simulations of halo stars, that is beyond the current study. 

\section{Summary}

We propose that investigators exploring increasingly more complex and sophisticated mass modeling of the Galaxy, utilize simple and robust mass estimates to constrain the range of mass priors that they explore. In this particular study, we use the timing argument and the H3 spectroscopic catalog to present such a limit. We use mock catalogs constructed from published, independent simulations to test our methodology and validate our conclusions. Our derived, conservative, 90\% confidence lower limit on the mass of the Milky Way, $0.91\times10^{12}$ M$_\odot$, already helps eliminate much of the mass range that previous models were unable to exclude.
Our analysis suggests a Milky Way mass of $\approx 1.4 \times 10^{12}$ M$_\odot$. The properties of stars at $R<60$ kpc qualitatively support these result, but those stars become increasingly more difficult to analyze as $R$ becomes smaller. Instead, we advocate for larger samples that may identify more stars at large $R$. 
Outer halo surveys are growing prodigiously. We expect a lower mass limit, such as that produced here, to continue to gain in confidence and be used generally to refine complex dynamical modeling.

\acknowledgements{We thank the Hectochelle operators Chun Ly, ShiAnne
Kattner, Perry Berlind, and Mike Calkins, and the CfA
and U. Arizona TACs for their continued support of this
long-term program. This research was supported in part
by the National Science Foundation under Grant No.
NSF PHY-1748958 and AST-1812461. DZ acknowledges the hospitality of
DARK, University of Copenhagen where the majority of this paper was written.
}

\facilities{MMT (Hectochelle)}

\begin{appendix}
\renewcommand\thefigure{\thesection.\arabic{figure}}  
\setcounter{figure}{0}
\section{R$_{200}$ vs. M$_{200}$}

\begin{figure}[t!]
\begin{center}
\includegraphics[width = 0.48 \textwidth]{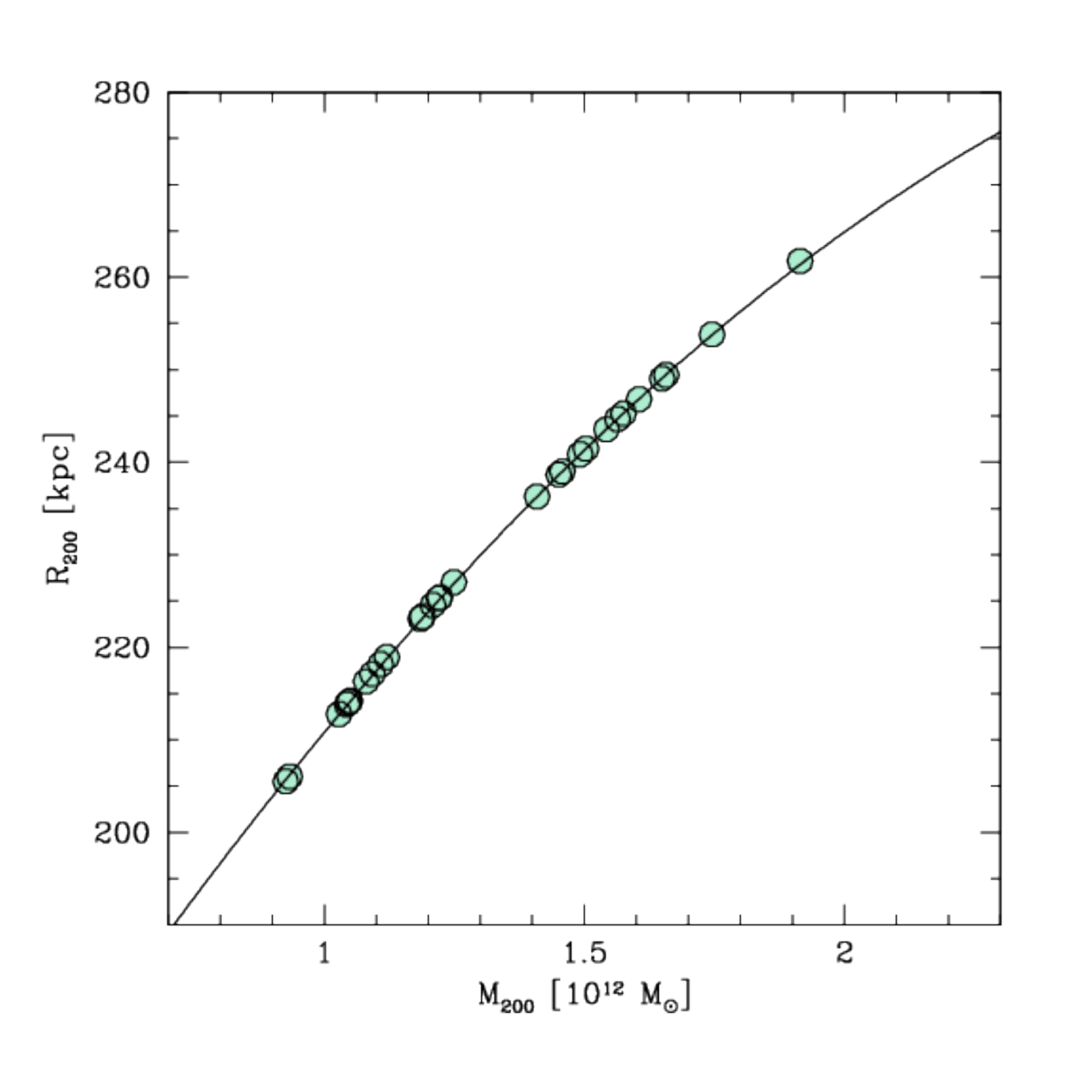}
\end{center}
\caption{R$_{200}$ vs. M$_{200}$ for simulated halos from the Auriga suite \citep{grand2019}. The curve is our low order fit to the relation that we use in estimating R$_{200}$ given M$_{200}$. The exact functional form is given in the text.}
\label{fig:scaling}
\end{figure}

When we rescale the calculated apocenter distances for the H3 stars, we require a calibration of M$_{200}$ vs R$_{200}$. In order to compare most directly to the Auriga simulations, we use the quoted values for all of the \cite{grand2017} models to derive an empirical relation. In Figure \ref{fig:scaling} we show both the data as presented by \cite{grand2017} and our low order fit to the relation given by
$${\rm R}_{200} = 128.968 + 95.7676 {\rm M}_{200} - 13.8829 {\rm M}_{200}^2,$$
where M$_{200}$ is in units of $10^{12}$ M$_\odot$ and R$_{200}$ in kpc.

\end{appendix}


\end{document}